\documentclass[12pt]{article}
\usepackage{amsfonts,amssymb,epsfig,amsmath}
\usepackage{color}

\usepackage[vcentermath]{youngtab}

\renewcommand{\baselinestretch}{1.2}

\begin{document}
%

\renewcommand{\theequation}{\thesection.\arabic{equation}}
\renewcommand{\thefootnote}{\alph{footnote}}

\begin{titlepage}

\begin{center}
\hfill {\tt SNUTP11-005}\\
\hfill {\tt KIAS-P11030}\\

\vspace{2cm}

{\large\bf Emergent Schr\"{o}dinger geometries from mass-deformed CFT}

\vspace{2cm}

\renewcommand{\thefootnote}{\alph{footnote}}

{
Hee-Cheol Kim$^1$, Seok Kim$^1$, Kimyeong Lee$^2$ and Jaemo Park$^3$}

\vspace{1cm}

\textit{$^1$Department of Physics and Astronomy \& Center for
Theoretical Physics,\\
Seoul National University, Seoul 151-747, Korea.}\\

\vspace{0.2cm}

\textit{ $^2$School of Physics, Korea Institute for Advanced Study,
Seoul 130-012, Korea.}\\

\vspace{0.2cm}

\textit{ $^3$Department of Physics \& Center for Theoretical Physics (PCTP),\\
POSTECH, Pohang 790-784, Korea.}\\

\vspace{0.7cm}


\end{center}

\vspace{1cm}

\begin{abstract}

We study an M-theory solution for the holographic flow of AdS$_4$ times
Sasaki-Einstein 7-manifolds with skew-whiffing, perturbed by a mass
operator. The infrared solution contains the 5 dimensional
Schr\"{o}dinger geometry after considering the gravity dual of the
standard non-relativistic limit of relativistic field theories.
The mass deformation of the field theory is discussed in detail
for the case with 7 manifold being a round sphere.

\end{abstract}

\end{titlepage}

\renewcommand{\thefootnote}{\arabic{footnote}}

\setcounter{footnote}{0}

\renewcommand{\baselinestretch}{1}

\tableofcontents

\renewcommand{\baselinestretch}{1.2}

\section{Introduction}

Gauge/gravity duality with non-relativistic scale invariance has drawn considerable
interest in recent years. An interesting direction of studies was based on the
Schr\"{o}dinger geometry \cite{Son:2008ye,Balasubramanian:2008dm}. See also \cite{Duval:1990hj}.
Schr\"{o}dinger symmetry is a scale invariant extension of the Galiean symmetry by
the particle number symmetry, anisotropic scale symmetry with dynamical exponent $z=2$
and special conformal symmetry. The geometry of \cite{Son:2008ye,Balasubramanian:2008dm}
realizes this symmetry of a QFT in $d+1$ spacetime dimension as isometries of a
$d+3$ dimensional spacetime of the following form\footnote{Such a geometry in lowest
possible dimension with $d\!=\!0$ has been studied in \cite{Israel:2004vv}, prior to
\cite{Son:2008ye,Balasubramanian:2008dm}.}:
\begin{equation}\label{schrodinger}
  ds^2=L^2\left[-\beta^2\rho^4(dx^+)^2+\rho^2(d\vec{x}^2+2dx^+dx^-)+\frac{d\rho^2}{\rho^2}
  \right]\ \ ,\ \ \ \vec{x}\in\mathbb{R}^d\ .
\end{equation}
In particular, the shift symmetry of the coordinate $x^-$ realizes
the particle number symmetry. There are many works on Schr\"{o}dinger solutions in
string/M-theory, including \cite{sch}, which also discuss Schr\"{o}dinger solutions with
general dynamical exponent $z$. See also many references below that we mention in more
detailed contexts.

$2+1$ dimensional non-relativistic CFT duals of 5 dimensional Schr\"{o}dinger geometries
(which we call Sch$_5$) were studied in \cite{Herzog:2008wg} with string theory embedding,
as deformations of $4$ dimensional relativistic field theories by irrelevant operators
which partly break the 4 dimensional conformal symmetry. See \cite{Guica:2010sw} for
more discussions on the holography in this context.

On the other hand, a very natural way of obtaining a non-relativistic theory
is to start from a relativistic CFT living in a spacetime with \textit{same dimension},
and to obtain the non-relativistic theory in the IR after a suitable mass deformation
and performing the standard non-relativistic limit of redefining the energy and discarding
`anti-particles.' Indeed, it has been shown concretely that a class of $2+1$ dimensional
mass-deformed CFT gives a non-relativistic theory with Schr\"{o}dinger symmetry
\cite{Nakayama:2009cz,Lee:2009mm}. Contrary to the other realization obtained by
starting from a relativistic CFT in one higher dimension, the Schr\"{o}dinger symmetry
cannot be embedded into the conformal symmetry of the relativistic theory.
It is thus interesting to see if there are mass-deformed CFT's admitting gravity duals
which show `holographic RG flows' from AdS$_D$ to Sch$_{D+1}$.\footnote{What
we mean by a flow solution from AdS$_D$ to Sch$_{D+1}$ will be clarified later, as the latter
lacks $D-1$ dimensional Poincare symmetry of AdS$_D$. We shall obtain Sch$_5$ after
combining an internal direction with four directions, performing the coordinate
transformation (\ref{coordinate-transform}) to have a time $x^+$ conjugate to the non-relativistic energy, and finally taking the scaling limit (\ref{scaling}).}
The case with $D\!=\!4$ will be our main interest in this paper, although similar studies
could be done in other dimensions.

In particular, \cite{Nakayama:2009cz,Lee:2009mm} discussed the non-relativistic superconformal
field theory obtained by deforming the $\mathcal{N}\!=\!6$ SCFT \cite{Aharony:2008ug} by
a mass term preserving all Poincare supersymmetries, considering a particular vacuum
in which all scalar expectation values are zero. The classical Lagrangian of this
non-relativistic field theory preserves $14$ supercharges, including $2$ extra non-relativistic
conformal supercharges. There have been some efforts to construct the gravity dual of this
theory in the IR based on the geometric realization of Schr\"{o}dinger symmetry \cite{Ooguri:2009cv,Jeong:2009aa}. None of them found it, at least with all $14$
Killing spinors. Soon it was argued \cite{Kim:2010mr,Cheon:2011gv} with a Witten index
calculation that the corresponding field theory vacuum does not exist,
at least in the subset of supersymmetric vacua. The gravity dual of this theory is still
ill-understood to date. See more comments about it in the discussion.

In this paper, we find a very simple class of flow solutions from AdS$_4$
to Sch$_5$ after deforming a class of relativistic CFT$_3$ with a mass operator.
The 3 dimensional relativistic CFT$_3$ can be regarded as those living on parallel M2-branes
in M-theory. Our construction is based on the relativistic UV CFT dual to the so-called skew-whiffed $AdS_4\times SE_7$, where $SE_7$ denotes 7 dimensional Sasaki-Einstein manifold.
The skew-whiffing in this context corresponds to changing the sign of M2-brane charge,
generically breaking supersymmetry. For the particular case of $SE_7=S^7$, maximal
superconformal symmetry is preserved and the UV CFT is relatively well understood.
In this case, we explicitly identify the mass operator, which turns out to be
non-supersymmetric and invariant under an $SU(4)\times U(1)$ subgroup of $SO(8)$.

Our finding shows how Schr\"{o}dinger geometries can appear by deforming
relativistic CFT's by mass operators and taking the non-realtivistic limit. It could find
applications in many different gauge/gravity dual models containing massive degrees
only. We briefly comment on some possible applications in the discussion.

The solution we consider in this paper can actually be understood as
the zero-temperature black brane solution found and discussed in \cite{Gauntlett:2009bh}
from 4 dimensional gauged supergravity. The 4 dimensional metric in the Einstein frame
in the `IR regime' (with small $r$) is given by
\begin{equation}
  ds^2\sim r^2(-dt^2+d\vec{x}^2)+\frac{dr^2}{r^{\frac{4}{3}}}\ \ ,
  \ \ \ \vec{x}\in\mathbb{R}^2
\end{equation}
which has zero entropy.
This type of geometry with Poincare symmetry and broken (isotropic) scale
invariance was considered in \cite{Charmousis:2010zz}, and a similar system was considered in
\cite{Iizuka:2011hg}. This type of geometry is different from those
considered in \cite{Goldstein:2009cv} which have approximate Lifshitz symmetry.
From our model, it is clear that the broken isotropic scale invariance
originates from the mass deformation of the UV CFT.

The remaining part of this paper is organized as follows. In section 2, we explain our
ansatz and obtain the analytic solution in the IR in the small radius expansion. We
explain the gravity dual operation of taking the non-relativistic limit of QFT, the
appearance of Sch$_5$ geometry, and the exact solution which interpolates our IR solution
with asymptotic AdS$_4$. In section 3, we explain the mass term of the dual field theory
when the internal 7-manifold is $S^7$ or orbifolds thereof. Section 4 concludes with
discussions.

\section{Flows from AdS to Schr\"{o}dinger geometry}

We seek for flow solutions from $AdS_4\times SE_7$ to 5 dimensional Schr\"{o}dinger
geometries in the consistent truncation ansatz discovered in \cite{Gauntlett:2009zw}.
It will turn out that we can realize a Schr\"{o}dinger solution in the infrared if we
advocate the so-called skew-whiffed
truncation. This truncated theory contains an $AdS_4\times SE_7$ vacuum, generically
with broken supersymmetry. For the special case of $SE_7=S^7$, the $AdS_4$ vacuum
is maximally supersymmetric. For the case with $\mathbb{Z}_k$ orbifold, the $AdS_4$ vacuum
after orbifold is non-supersymmetric with skew-whiffing.
The nature of the mass term driving the flow of our solution will be discussed in section 3 after
studying the gravity solutions. The full solution is actually available in an
analytic form \cite{Gauntlett:2009bh}, but to illustrate how the solution looks like
and to explain why skew-whiffed setting is necessary to have nontrivial solutions,
we explain some details of the derivation.

The consistent truncation ansatz of \cite{Gauntlett:2009zw} is
\begin{eqnarray}\label{truncation}
  ds_{11}^2&=&ds_4^2+e^{2U}ds^2(KE_6)+e^{2V}(\eta+A_1)^2\nonumber\\
  G_4&=&6e^{-6U-V}(\epsilon+h^2+|\chi|^2){\rm vol}_4+H_3\wedge (\eta+A_1)+H_2\wedge J
  +dh\wedge J\wedge (\eta+A_1)+2hJ\wedge J\nonumber\\
  &&+\sqrt{3}\left[\chi(\eta+A_1)\wedge\Omega-\frac{i}{4}D\chi\wedge\Omega+c.c.\right]\ ,
\end{eqnarray}
where $J$ is the K\"{a}hler 2-form of the K\"{a}hler-Einstein base $KE_6$, $\eta=d\psi+\theta$ is
the Reeb 1-form with $d\theta=2J$, $D\chi\equiv d\chi-4iA_1\chi$. $A_1$, $H_2$, $H_3$ are
1-form, 2-form, 3-form in 4 dimensions, while $U,V,h$ and $\chi$ are real/complex
scalars, respsectively. The truncation with $\epsilon=\pm 1$ contains supersymmetric and generically non-supersymmetric $AdS_4$ vacuum. The case with $\epsilon=-1$ is the skew-whiffed truncation.

We would like to study the flow solutions preserving the $2+1$ dimensional Poincare symmetry
on the M2-brane world-volume. This is because the flow itself after a mass deformation does not
break Poincare symmetry on the worldvolume. It is only the non-relativistic limit, discarding
`anti-particles' after suitably redefining non-relativistic energy, which breaks it.
We thus consider the Poincare invariant ansatz
\begin{equation}\label{4d-metric}
  ds_4^2=e^{2A(r)}(\eta_{\mu\nu}dx^\mu dx^\nu)+dr^2\ .
\end{equation}
From Poincare symmetry, the gauge fields $H_2$, $F_2$ are zero. At this stage,
it appears that $H_3$ may be nonzero if proportional to the volume form
$dx^0\wedge dx^1\wedge dx^2$.
From the equations of motion and the Bianchi identities for $G_4$, one obtains
\begin{equation}
  dH_3=0\ ,\ \ dH_2=2H_3+H_2\wedge F_2\ ,\ \ H_1=dh
\end{equation}
and
\begin{eqnarray}
  d(e^{6U-V}\star H_3)&=&e^{6U+V}fF_2-6e^{2U+V}\star H_2-12hH_2-\frac{3i}{2}D\chi\wedge
  D\chi^\ast\nonumber\\
  d(e^{2U+V}\star H_2)&=&-2dh\wedge H_2-4hH_3\nonumber\\
  d(e^{2U-V}\star dh)&=&-e^{2U+V}\star H_2\wedge F_2-H_2\wedge H_2-4h(f+4e^{-2U+V}){\rm vol}_4\nonumber\\
  D(e^V\star D\chi)&=&-iH_3\wedge D\chi-4\chi(f+4e^{-V}){\rm vol}_4\nonumber\\
  f&=&6e^{-6U-V}(\epsilon+h^2+|\chi|^2)\ .
\end{eqnarray}
From the $H_2$ equation of motion on the second line,
one should set $hH_3=0$. So either $h$ or $H_3$ should
be vanishing. As we shall see later, nonzero $h$ is crucial to have a Schr\"{o}dinger
solution in the IR, similar to those in \cite{O'Colgain:2009yd}. So we set $H_3=0$.
The field $\chi$ can always be consistently set to zero, which we do from now on.
Then, only the $h$ equation of motion (third line)
remains nontrivial among the above equations.

As we shall explain shortly, we want an IR solution with $U=const$,
$e^V\sim e^A\rightarrow 0$ as $r\rightarrow-\infty$, to get 5 dimensional
Schr\"{o}dinger solutions. It turns out that one can set $U=0$, which is the value for
the AdS$_4$ vacuum, all along the flow. We shall later explain that this is a consistent
ansatz, which actually comes from the existence of a further consistent truncation found
in \cite{Gauntlett:2009bh}. With $U\!=\!0$, the 11 dimensional Einstein equation becomes
\begin{eqnarray}
  R_{mn}&=&\nabla_m\nabla_n V+\nabla_m V\nabla_n V
  +\frac{3}{2}e^{-2V}(\nabla_m h\nabla_n h-\frac{1}{3}g_{mn}(\nabla h)^2)
  -2g_{mn}(4h^2+\frac{f^2}{6})\nonumber\\
  0&=&\nabla^m\nabla_m V+\nabla^m V\nabla_m V+e^{-2V}\nabla^m h\nabla_m h
  -6e^{2V}-8h^2+\frac{f^2}{6}\nonumber\\
  0&=&-8+2e^{2V}+8h^2+\frac{f^2}{6}
\end{eqnarray}
with $m,n=0,1,2,r$. The last equation is the $U$ equation of motion.
Note that $f=6e^{-V}(\epsilon+h^2)$. With our 4 dimenisional metric (\ref{4d-metric}),
one obtains
\begin{equation}
  R_{rr}=-3(A^{\prime\prime}+(A^\prime)^2)\ ,\ \ R_{\mu\nu}=-(A^{\prime\prime}+3(A^\prime)^2)
  g_{\mu\nu}\ ,\ \ \Gamma_{\mu\nu}^{\ \ r}=-A^\prime g_{\mu\nu}
\end{equation}
for $\mu,\nu=0,1,2$, where prime denotes $r$ derivative.
Therefore, the two nontrivial components ($rr$ and $\mu\nu$) of the
Einstein equations and the $V,U,h$ equations become
\begin{eqnarray}\label{5-equations}
  0&=&3(A^{\prime\prime}+(A^\prime)^2)+V^{\prime\prime}+(V^\prime)^2+e^{-2V}(h^\prime)^2
  -8h^2-12e^{-2V}(\epsilon+h^2)^2\nonumber\\
  0&=&A^{\prime\prime}+3(A^\prime)^2+A^\prime V^\prime-\frac{1}{2}e^{-2V}(h^\prime)^2-8h^2-12e^{-2V}(\epsilon+h^2)^2\nonumber\\
  0&=&V^{\prime\prime}+3A^\prime V^\prime+(V^\prime)^2+e^{-2V}(h^\prime)^2-6e^{2V}-8h^2
  +6e^{-2V}(\epsilon+h^2)^2\nonumber\\
  0&=&-8+2e^{2V}+8h^2+6e^{-2V}(\epsilon+h^2)^2\nonumber\\
  0&=&h^{\prime\prime}+(3A^\prime-V^\prime)h^\prime-4he^V[4e^V+6e^{-V}(\epsilon+h^2)]\ .
\end{eqnarray}
Apparently, there are 5 equations for 3 functions $A,V,h$. One more equation appears
due to our Poincare invariant ansatz with $r$ dependent functions only, and another one
due to our ansatz $U\!=\!0$. The former extra equation will be eliminated from the
consistency of Poincare invariant ansatz, and the second one from the consistency
of $U=0$ ansatz.

We first explain some structures of these equations. As briefly explained above,
we want $e^V\sim e^A\sim e^{nr}\rightarrow 0$ for some constant $n$ as $r\rightarrow-\infty$.
Then the terms containing $A^\prime$ and $V^\prime$ in the first three equations all yield $4n^2$,
which are of order $1$. Also, the last factors $\sim e^{-2V}(\epsilon+h^2)^2$ become
very large due to $e^{-2V}$, unless $\epsilon+h^2$ is small. Especially, in the second equation,
considering the signs of other terms, this term cannot be canceled at all if it is indeed large.
The only way to have the desired IR solution is to take $\epsilon\!=\!-1$ with skew-whiffing and
demand $h\rightarrow\pm 1$ as $r\rightarrow-\infty$. In section 3, we will also provide a group
theoretical argument from UV perspective on why the desired flow solutions exist only with
skew-whiffing in this truncation.

Firstly, the equation for $U$ becomes to
\begin{equation}
 0=-8+2e^{2V}+8h^2+6e^{-2V}-12h^2e^{-2V}+6e^{-2V}h^4=
 2[e^V+3(-1+h^2)e^{-V}][e^V+(-1+h^2)e^{-V}]\ .
\end{equation}
This relates $h$ and $e^V$ in an algebraic manner. It will turn out that
\begin{equation}\label{h-V}
  e^{2V}=1-h^2
\end{equation}
is the solution that we want. It is a special case of an algebraic
relation found in the truncation of \cite{Gauntlett:2009bh} with $\chi=0$.
We are left with 4 equations for 2 variables $A,V$. One can easily show that the $h$
equation is equivalent to the $V$ equation. To show this, we multiply $h$
to the last equation of (\ref{5-equations}) and obtain
\begin{equation}
  0=(hh^\prime)^\prime-(h^\prime)^2+\frac{1}{2}(3A^\prime-V^\prime)(h^2)^\prime+8h^2e^{2V}
  =-e^{2V}[V^{\prime\prime}+2(V^\prime)^2+e^{-2V}(h^\prime)^2+(3A^\prime-V^\prime)
  V^\prime-8h^2]
\end{equation}
after inserting (\ref{h-V}). As the expression in the last parenthesis is the $V$ equation,
the two equations are equivalent as long as $h\neq 0$. In our flow solution, $h$
will only be asymptotically zero in UV, where we can easily check separately that the UV
fixed point solves all equations. So the last constraint will not obstruct
the above elimination of one equation.

So far, the three equations for the two variables $A,V$ (with (\ref{h-V}) assumed) are
\begin{eqnarray}\label{3-eqns}
  0&=&3(A^{\prime\prime}+(A^\prime)^2)+V^{\prime\prime}+(V^\prime)^2+e^{-2V}(h^\prime)^2
  -8-4e^{2V}\nonumber\\
  0&=&A^{\prime\prime}+3(A^\prime)^2+A^\prime V^\prime
  -\frac{1}{2}e^{-2V}(h^\prime)^2-8-4e^{2V}\nonumber\\
  0&=&V^{\prime\prime}+(V^\prime)^2+3A^\prime V^\prime
  +e^{-2V}(h^\prime)^2-8(1-e^{2V})\ ,
\end{eqnarray}
where the last equation may be replaced by
\begin{equation}\label{h-equation}
  0=h^{\prime\prime}+(3A^\prime-V^\prime)h^\prime+8he^{2V}=
  -h^{-1}e^{2V}\times [V]\ ,
\end{equation}
where $[V]$ denotes the expression in the $V$ equation of motion,
on the last line of (\ref{3-eqns}). Also, we denote by $[R_{rr}]$ and $[R_{\mu\nu}]$
the expressions on the first and second lines of (\ref{3-eqns}).
To show that one of these three equations is redundant, we find it useful
to consider the following two combinations of three independent equations:
\begin{eqnarray}\label{basis}
  \frac{1}{3}[R_{rr}]-\frac{1}{3}[V]&:&0=A^{\prime\prime}+(A^\prime)^2-A^\prime V^\prime-4e^{2V}\\
  \left[R_{rr}\right]-[R_{\mu\nu}]&:&2A^{\prime\prime}+V^{\prime\prime}+(V^\prime)^2
  -A^\prime V^\prime+\frac{3}{2}e^{-2V}(h^\prime)^2\ .\nonumber
\end{eqnarray}
Let us take 3 combinations of equations to be $\frac{[R_{rr}]-[V]}{3}=0$,
$[R_{\mu\nu}]=0$ and
\begin{eqnarray}
  0&=&h^\prime\left(-he^{2V}[V]\right)-2e^{2V}A^\prime\left([R_{rr}]-[R_{\mu\nu}]\right)
  -2e^{2V}V^\prime\left(\frac{1}{3}[R_{rr}]-\frac{1}{3}[V]\right)\nonumber\\
  &=&e^{2V}\left[\frac{1}{2}e^{-2V}(h^\prime)^2-2A^\prime V^\prime-2(A^\prime)^2\right]^\prime.
\end{eqnarray}
The last total derivative can be obtained after some straightforward algebra.
So this equation can be integrated to yield
\begin{equation}
  0=\frac{1}{2}e^{-2V}(h^\prime)^2-2A^\prime V^\prime-2(A^\prime)^2-c\nonumber\\
  =\left(\frac{1}{3}[R_{rr}]-\frac{1}{3}[V]\right)-[R_{\mu\nu}]\ ,
\end{equation}
where the last equation holds for $c=-8$. Thus, the first two equations guarantee this
last equation with the integration constant chosen as $c=-8$. So we have
two independent equations for two functions in our ansatz.

To get IR solutions near $r\!=\!-\infty$, we define $\rho\equiv e^{nr}$ with
a positive number $n$ and expand
\begin{equation}
  h=1+h_1\rho^2+\cdots\ ,\ \ e^{V}=v_0(\rho+v_1\rho^3+\cdots)\ ,\ \
  e^A=a_0(\rho+a_1\rho^3+\cdots)\ .
\end{equation}
We can set $v_0\!=\!1$ without losing generality by shifting the variable $r$,
as all equations have translation symmetry in $r$. $a_0$ can also be eliminated to,
say $a_0=1$, by rescaling the Poincare coordinates $x^\mu$. We leave it as it is
as we want to identify it with a mass parameter in the UV later.
Plugging these in, we find $n^2\!=\!2$ at $\mathcal{O}(\rho^0)$
terms of the equations. Continuing the iteration, one finds the following
small radius expansion in $\rho=e^{\sqrt{2}r}$
\begin{eqnarray}\label{small-radius}
  e^V&=&\rho\left(1-\frac{7}{16}\rho^2+\frac{69}{256}\rho^4-\frac{2209}{12288}\rho^6
  +\cdots\right)\nonumber\\
  e^A&=&a_0\rho\left(1+\frac{3}{16}\rho^2-\frac{15}{256}\rho^4+\frac{329}{12288}\rho^6
  +\cdots\right)\ ,
\end{eqnarray}
with $U=0$ and $h^2=1-e^{2V}$.

The solution with leading and next-to-leading terms is given by
\begin{eqnarray}\label{leading}
  ds_{11}^2&=&a_0^2\rho^2\left(1+\frac{3}{16}\rho^2+\mathcal{O}(\rho^4)\right)^2
  (-dt^2+d\vec{x}^2)+\frac{d\rho^2}{2\rho^2}+
  \rho^2\left(1-\frac{7}{16}\rho^2+\mathcal{O}(\rho^4)\right)^2\eta^2
  +ds^2(KE_6)\nonumber\\
  G_4&=&-6a_0^3\rho^4\left(1+\mathcal{O}(\rho^2)\right){\rm vol}_3\wedge dr
  -\frac{1}{2}d\left(\rho^2+\mathcal{O}(\rho^4)\right)\wedge J\wedge\eta
  +\left(2+\mathcal{O}(\rho^2)\right)J\wedge J\ ,
\end{eqnarray}
where 
vol$_3=dt\wedge dx^1\wedge dx^2$.
To obtain the Schr\"{o}dinger solution, we first perform a coordinate transformation
to a rotating frame. Taking $\eta=d\psi+\theta$, where $d\theta=2J$ is
twice the K\"{a}hler 2-form of $KE_6$, we define
\begin{equation}\label{coordinate-transform}
  x^-=\psi-a_0t\ ,\ \ x^+=t
\end{equation}
to be the coordinates in the rotating (or boosted) frame.
For later use, we note that $\psi$ is $2\pi$ periodic for $SE_7\!=\!S^7$
in our normalization. From the dual field theory
perspective, the shift of $\psi$ is identified as the particle number symmetry, or more
precisely the symmetry conjugate the total rest mass as there could be more than one
particle species. So the conserved charge for $x^+$ takes the form of the non-relativistic
energy $E_{NR}=E_{rel}-a_0R_b$, where $E_{rel}$ and $R_b$ are conserved charges for the
translation of $t,\psi$. The above metric can be written as
\begin{eqnarray}
  ds_{11}^2&=&-a_0^2\rho^2\left(\frac{5}{4}\rho^2+\mathcal{O}(\rho^4)\right)(dx^+)^2+
  a_0^2\rho^2\left(1+\mathcal{O}(\rho^2)\right)^2d\vec{x}^2+
  2a_0\rho^2(1+\mathcal{O}(\rho^2))dx^+\eta\nonumber\\
  &&+\frac{d\rho^2}{2\rho^2}+\rho^2\left(1+\mathcal{O}(\rho^2)\right)\eta^2
  +ds^2(KE_6)\ .
\end{eqnarray}
Note that $\rho^2(dx^+)^2$ terms cancels by the coordinate transformation,
leaving the leading $\rho^4(dx^+)^2$ term.
Taking the scaling limit
\begin{equation}\label{scaling}
  \rho\rightarrow\epsilon \rho\ ,\ \ x^+\rightarrow\epsilon^{-2}x^+\ ,\ \
  \vec{x}\rightarrow\epsilon^{-1}\vec{x}\ ,\ \ x^-\rightarrow x^-
\end{equation}
with $\epsilon\rightarrow 0^+$, one obtains
\begin{eqnarray}\label{IR-solution}
  ds_{11}^2&=&-\frac{5a_0^2}{4}\rho^4(dx^+)^2+\rho^2\left(a_0^2d\vec{x}^2+
  2a_0dx^+(dx^-+\theta)\right)+\frac{d\rho^2}{2\rho^2}+ds^2(KE_6)\nonumber\\
  G_4&=&-3\sqrt{2}a_0^3\rho^3dx^+\wedge dx^1\wedge dx^2\wedge d\rho
  -a_0\rho d\rho\wedge J\wedge dx^++2J\wedge J\ .
\end{eqnarray}
The 5 dimensional metric apart from the $KE_6$ part is the Schr\"{o}dinger geometry
dual to $2+1$ dimensional non-relativistic field theory. The coordinate transformation
and scaling explained above are to be understood as the gravity dual of the
non-relativistic limit.

Sch$_5$ solutions in M-theory of this type was found in \cite{O'Colgain:2009yd}. However,
the solution found there has different coefficients in various terms.
In fact, one can find a generalized solution
\begin{eqnarray}\label{generalized}
  ds_{11}^2&=&-\frac{14a_0^2-b_0^2}{8}\rho^4(dx^+)^2+\rho^2\left(a_0^2d\vec{x}^2+
  b_0dx^+(dx^-+\theta)\right)+\frac{d\rho^2}{2\rho^2}+ds^2(KE_6)\nonumber\\
  G_4&=&-3\sqrt{2}a_0^3\rho^3dx^+\wedge dx^1\wedge dx^2\wedge d\rho
  -a_0\rho d\rho\wedge J\wedge dx^++2J\wedge J\ .
\end{eqnarray}
which includes our solution (\ref{IR-solution}) and that of \cite{O'Colgain:2009yd}
as special cases. As before, one can eliminate one of the two parameters $a_0,b_0$
by scaling $x^+,\vec{x}$ in same ratio, after which one parameter remains. Setting
$b_0=2a_0$ yields our solution (\ref{IR-solution}), while setting $b_0=a_0$ yields
that of \cite{O'Colgain:2009yd}.\footnote{We find that their eqn.(4.4) becomes
a solution if we take $f_0=\frac{13\alpha^2}{64}$ (with $c=2$ which is our normalization),
correcting a minor typo in their expression $f_0=\frac{13\alpha}{4c^4}$. Relating their
quantities with ours as $r_{\rm theirs}=\sqrt{2}a_0\rho_{\rm ours}$, $x^-_{\rm theirs}=
a_0^{-1}x^-_{\rm ours}$, $\alpha=-2/a_0$, $C_1=-a_0^{-1}\theta$ with same $x^+$, their
eqn.(4.4) takes the form of (\ref{generalized}) above.}

The `leading behavior' of the IR solution (\ref{leading}) before
coordinate redefinition, keeping $e^A,e^V\sim\rho$ only, might look like
$AdS_5$ fibered over $KE_6$ with a compact longitudinal direction along $\psi$.
However, this $AdS_5$ fibered over $KE_6$ \textit{does not} satisfy the equations of
motion (\ref{5-equations}) itself, although making the leading $\mathcal{O}(\rho^0)$
equations satisfied. Namely, there are subleading terms in the equations
caused by the leading terms of the functions $e^A,e^V$, to be canceled by the subleading terms
of the functions. There exist $AdS_5\times KE_6$ solutions in M-theory without any
fibration in the $\psi$ angle, which is related to our `leading behavior' by replacing $\eta\rightarrow d\psi$ with same radii for $AdS_5$ and $KE_6$ \cite{Dolan:1984hv}.
Another point to mention is that this $AdS_5$ cannot be obtained by a scaling limit
similar to the one we explained above for $Sch_5$. Naively, it might appear that
the scaling $\rho\rightarrow\epsilon\rho$, $t\rightarrow\epsilon^{-1}t$,
$\vec{x}\rightarrow\epsilon^{-1}\vec{x}$, $\psi\rightarrow\epsilon^{-1}\psi$ with
$\epsilon\rightarrow 0$ would do this job. However, the last $\psi$ direction is compact.
This scaling limit demands one to zoom out to the long wavelength along
$t,\vec{x},\psi$, which does not make sense with a compact direction with a maximal
wavelength. Note that the scaling (\ref{scaling}) to the Schr\"{o}dinger geometry
leaves $x^-$ unchanged.

Before studying the full flow solution which interpolates the above IR solution with
AdS$_4$ UV fixed point, we review the asymptotic expansion in UV \cite{Gauntlett:2009bh}.
We obtain
\begin{equation}
  e^A=e^{2r}\left(1+a_1e^{-nr}+\cdots\right)\ ,\ \ e^V=1-v_1e^{-nr}+\cdots\ ,\ \
  h^2=2v_1e^{-nr}+\cdots\ ,
\end{equation}
with two possible solutions for the coefficients
\begin{equation}\label{UV-modes}
  n=4\ ,\ \ v_1=\frac{8}{3}a_1\ \ \ {\rm and}\ \ \ n=8\ ,\ \ v_1=2a_1\ .
\end{equation}
Defining $z\equiv e^{-2r}$, one obtains the asymptotic AdS$_4$ metric
$ds_4^2=\frac{d\tilde{x}_\mu d\tilde{x}^\mu+dz^2}{4z^2}$ upon defining
$\tilde{x}^{\mu}=2x^\mu$. The general asymptotic solution is
\begin{equation}
  h\sim h_1z^{\Delta_-}+h_2z^{\Delta_+}+\cdots
\end{equation}
with $\Delta_-\!=\!1$ and $\Delta_+\!=\!2$. Both modes are normalizable with these
coefficients and there are two possible dual CFT's in which one mode is the source
of a boundary operator whose expectation value is given by the coefficient
of the other mode. We stick to the CFT in which the field $h$ sources the fermion mass
term at the boundary as we review in detail in section 3, in which the mode with
$\Delta_-=1$ is taken to be the source. The fermion mass operator comes
with dimension $\Delta_+=2$. From the gravity solution, one can get the precise values
of the fermion and boson masses, which we explain in section 3.

\begin{figure}[t!]
  \begin{center}
    \includegraphics[width=8cm]{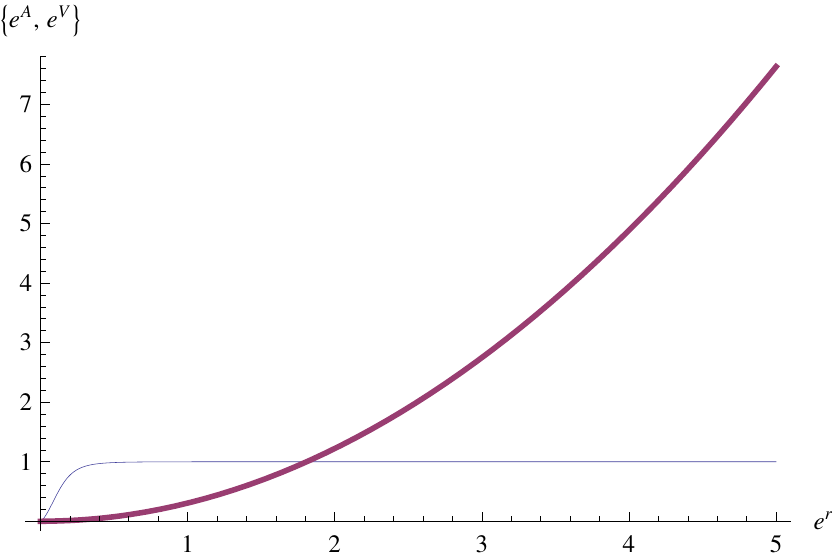}
\caption{Plot of the numerical solutions as functions of
$e^r$. The thick/thin lines are plots of $e^A$ and $e^V$, respectively. $e^V$ approaches
$1$ as $r\rightarrow\infty$ (the value for the $AdS_4$ vacuum).}\label{flow}
  \end{center}
\end{figure}
We now study if the IR solution can be interpolated to the asymptotic AdS$_4$
solution at $r=\infty$. We chose the first equation of (\ref{basis}) and the $h$ equation
(\ref{h-equation}) for numerics. The numerical analysis of the differential equations
exhibits the desired interpolation. The result for $a_0=.01$ is shown in Fig \ref{flow}.
One can also vary the values of $a_0$ and numerically relate it with the variable $v_1$
in the UV expansion. In doing so, we should remember that both in the IR and UV solutions
we implicitly tuned a coefficient by using the translational symmetry of $r$ in the equations.
In matching the two solutions, only one tuning is allowed. So we relax our UV asymptotic
behavior by admitting the $r$ translation degrees and write the solution as
\begin{equation}\label{UV-coefficients}
  e^A=ce^{2r}+a_1c^{-1}e^{-2r}+\cdots\ ,
\end{equation}
etc., where $a_1$ is the same coefficient as above. We fit the two coefficients $a_1,c$ by
comparing the expansion (\ref{UV-coefficients}) with the numerical solution at two large
values of $r$, which we chose as $e^r=20,100$, and find $c$, $a_1=\frac{3v_1}{8}$ as
functions of $a_0$. We find the relation to be
\begin{equation}\label{UV-IR-relation}
  a_0=\sqrt{v_1}\ \ ,\ \ \ c=a_0\ .
\end{equation}
$\sqrt{v_1}$ appearing as the leading coefficient of $h$ in UV is the mass for fermions,
consistent with the IR expectation from (\ref{coordinate-transform}) that $a_0$ should
be proportional to the mass parameter.

Surprisingly, one can find an exact analytic solution for the whole flow
\cite{Gauntlett:2009bh},
\begin{eqnarray}
  ds_{4E}^2&=&=r^2(-dt^2+d\vec{x}^2)+g^{-1}(r)dr^2\\
  h(r)&=&\frac{\sqrt{3}\sqrt{\frac{6\sqrt{3}(2r^8+4r^4+1)}{\sqrt{Z}}-Z+9}-\sqrt{3Z}+3}
  {6(r^4+1)}\nonumber\\
  Z(r)&\equiv&\frac{2\cdot 6^{2/3}(r^4+1)r^{8/3}}{(\sqrt{48r^4+81}-9)^{1/3}}-6^{1/3}
  (r^4+1)\left(\sqrt{48r^4+81}-9\right)^{1/3}r^{4/3}+3\nonumber\\
  g^{-1}(r)&=&\frac{\sqrt{1-h^2}}{4r^2}\left(1-\frac{r^2(h^\prime)^2}{4(1-h^2)^2}\right)
  \ ,\nonumber
\end{eqnarray}
where the subscript $E$ denotes the 4 dimensional metric in the Einstein frame.
(Note that their $r$ above is not our $r$ used previously.) It is related to our
4 dimensional metric as $(g_E)_{\mu\nu}=e^Vg_{\mu\nu}$.
We managed to check this exact solution by a heavy use of mathematica.
We also checked that our IR expansion (\ref{small-radius}) is completely reproduced
from the above exact solution, after relating our IR radial variable $\rho$ with their $r$
as
\begin{equation}
  \rho=2^{1/3}\left(r^{2/3}+\frac{r^2}{24\cdot 2^{1/3}}-\frac{7r^{10/3}}{288\cdot 2^{2/3}}
  +\frac{767r^{14/3}}{82944}+\cdots\right)\ ,
\end{equation}
according to $e^{V/2}\frac{d\rho}{2^{1/2}\rho}=\frac{dr}{g^{1/2}}$.

\section{Mass deformation of the UV CFT}

When the 7 manifold is $SE_7=S^7$ or its $\mathbb{Z}_k$ orbifolds, we can identify the
mass term which drives our flow. At $k=1$, the M2-brane worldvolume scalars and
fermions transform under $SO(8)$ as a vector ${\bf 8}_v$ and a chiral spinor ${\bf 8}_s$,
respectively. The $\mathcal{N}\!=\!8$ supercharge as well as
the conformal supercharge transform as ${\bf 8}_c$. Our skew-whiffed truncation ansatz
has a preferred $SU(4)\times U(1)_b$ symmetry \cite{Imaanpur:2010yk,Forcella:2011pp}:
in particular, the ansatz is invariant under $SU(4)$ isometry of the $KE_6=\mathbb{CP}^3$
base. The three $SO(8)$ representations decompose under $SU(4)\times U(1)_b$ as
\begin{equation} \label{skew-whiffed-rep}
  {\bf 8}_v\rightarrow {\bf 4}_{-1}+\bar{\bf 4}_{1}\ ,\ \
  {\bf 8}_s\rightarrow {\bf 1}_2+{\bf 6}_0+{\bf 1}_{-2}\ ,\ \
  {\bf 8}_c\rightarrow {\bf 4}_1+\bar{\bf 4}_{-1}\ .
\end{equation}
The charges of $U(1)_b$ are chosen so that $Q({\rm scalar})=({\rm fermion})$ is possible,
correcting a minor sign error in \cite{Forcella:2011pp}. The M2-brane theory without
skew-whiffing can be obtained by a parity transformation on $\mathbb{R}^8$, interchanging
${\bf 8}_s$ and ${\bf 8}_c$ representations. After this change, the supercharges
in ${\bf 6}_0$ of ${\bf 8}_s$ are the $\mathcal{N}\!=\!6$ supersymmetries of
\cite{Aharony:2008ug}. With skew-whiffing, all the supercharges transforming in ${\bf 4}_1+\bar{\bf 4}_{-1}$ are broken for $k>1$. At $k=1$, monopole operators with $U(1)_b$
charge $\pm 1$ make them gauge-invariant \cite{Kim:2009wb}, enabling maximal supersymmetry.

We would like to explain the nature of the mass term triggering our flow.
The mass term should be invariant under the $SU(4)$ symmetry of the $\mathbb{CP}^3$ in our
truncation, in the skew-whiffed frame. Decomposing ${\bf 8}_s$ (matter fermions)
into ${\bf 1}_2+{\bf 6}_0+{\bf 1}_{-2}$, we call them $\xi$, $\Psi^I$, $\bar\xi$
with $I=1,2,\cdots,6$, following \cite{Forcella:2011pp}. The fermions $\Psi^I$
($I=1,2,\cdots,6$) in ${\bf 6}_0$ are Majorana spinors, proposed to be in the
adjoint representation of the first $U(N)$, while the complex fermion $\xi$ in
${\bf 1}_2$ is in the bi-fundamental representation of $U(N)\times U(N)$. It seems that
the parity symmetry of this system is not manifest in the field theory.
In our discussions below, the role of these gauge group representations will be rather
minor. The mass for real fermions $\psi^a$, with $a=1,2,\cdots,8$ in ${\bf 8}_s$
of $SO(8)$, is given by (at least on a single M2-brane)
\begin{equation}
  M_{ab}\psi^a\psi^b
\end{equation}
with an $8\times 8$ symmetric mass matrix $M_{ab}$.
The symmetric matrix has $36$ real components and decomposes under $SO(8)$ as
${\bf 35}_++{\bf 1}$, where ${\bf 35}_+$ is the `traceless' part with respect to the
charge conjugation matrix, and ${\bf 1}$ is the `identity' part.
The spinor bi-linear in ${\bf 35}_+$ of $SO(8)$ is equivalent to self-dual
4-forms in $\mathbb{R}^8$, and will turn out to be dual to the self-dual 4-form modes in
the gravity dual \cite{Bena:2000zb} as we explain in some detail below. For now, we just
mention that the last two terms on the second line of (\ref{truncation}) with nonzero $h$
provides the relevant 4-form flux.

The mass matrix decomposes under $SU(4)$ as \cite{Imaanpur:2010yk}
\begin{equation}
  {\bf 35}_+\rightarrow {\bf 1}_0+{\bf 1}_4+{\bf 1}_{-4}+{\bf 6}_2+{\bf 6}_{-2}+{\bf 20}_0\ ,
\end{equation}
More concretely, the possible mass terms (on a single M2-brane) look like
\begin{equation}
  \frac{m_1}{2}\Psi^I\Psi^I+m_1^\prime\bar\xi\xi+
  {\rm Re}\left[m_2\xi^2+m_I\Psi^I\xi\right]+\frac{m_{IJ}}{2}\Psi^I\Psi^J
\end{equation}
with real masses $m_1$ and $m_1^\prime$, real traceless symmetric mass matrix $m_{IJ}$, complex
masses $m_2$, $m_I$. All but one combination of the first two mass terms proportional to
$m_1$, $m_1^\prime$ belong to ${\bf 35}_+$. A combination of the first two becomes
an $SO(8)$ invariant. This cannot be dual to the 4-form flux that we turn on in our solution, as
4-form fluxes in $\mathbb{R}^8$ always break $SO(8)$. So we do not consider this
combination. The other `traceless' combination $m_1^\prime=-3m_1$, with canonically normalized
kinetic terms for the fermions, corresponds to ${\bf 1}_0$ in the above decomposition of
${\bf 35}_+$. This is our mass term,
as our gravity solution is invariant under $SU(4)\times U(1)_b$.
For multiple M2-branes, the first two mass terms proportional to $m_1,m_1^\prime$ can be
made gauge invariant by taking traces, as $\Psi^I$ is an adjoint and
$\xi,\bar\xi$ are in conjugate representations.
We emphasize that, as fermions are in ${\bf 8}_c$ without skew-whiffing, the mass matrix,
now given by anti-self-dual 4-forms, decomposes under $SU(4)$ as ${\bf 35}_-\rightarrow
{\bf 10}+{\bf 10}+{\bf 15}$ without singlets. This explains why our consistent truncation
only has solutions with skew-whiffing.

The bosonic mass term is also $SU(4)$ invariant, which is
unique up to coefficient:
\begin{equation}
  m_b\bar{Z}_iZ^i
\end{equation}
with $i=1,2,3,4$. This mass term is actually invariant under $SO(8)$, at least
for the single M2-brane for which we can decompose complex scalars to real scalars
in ${\bf 8}_v$. Even at $k=1$ in which the UV CFT is supersymmetric, it is easy to
see that the mass-deformed theory is always non-supersymmetric.
This is simply because the mass is same for all bosonic fields, while there appear
two different nonzero masses for fermions.

One can compute the precise values of the fermion/boson masses
with the known form of the flux. Since we have a CFT before mass deformation,
physically meaningful quantity is just the ratio $m_1/m_b$ of the two masses.
To conveniently match factors, it is useful to compare the normalization with
the well known maximally supersymmetric mass deformation.
As a warming up, let us first consider an illustrative example of single M2-brane
coupled to a background 4-form field $G_4=dC_3$ in (almost) flat space. We
shall later explain that the same result is obtained from the large $N$ calculation
in $AdS_4\times S^7$. The coupling to this field is given by
\begin{equation}\label{wv-fermion-mass}
  \int_{\mathbb{R}^{2,1}}\alpha[C_3]_{012}++\beta\bar\psi\gamma^{IJKL}\psi[G_4]_{IJKL}
  +\cdots\ ,\ \ \ \ (I,J,K,L=1,2,\cdots,8)
\end{equation}
with coefficients $\alpha,\beta$ determined in \cite{Lambert:2009qw}.
We turn on small constant self-dual $G_4$ on $\mathbb{R}^8$,
where smallness is to keep the back reaction to the flat space to be small.
The bosonic mass $m_b$ is related to the constant flux as \cite{Lambert:2009qw}
\begin{equation}\label{boson-mass}
  m_b^{\ 2}=\frac{1}{8\cdot 4!}G^2\ .
\end{equation}
This comes partly from the first term of (\ref{wv-fermion-mass})
after the constant self-dual 4-form back-reacts to $(C_3)_{012}$, and also partly
from the back-reaction to the metric. For the two cases
\begin{equation}
  G_4=\gamma_0(dx^1\wedge dx^2\wedge dx^3\wedge dx^4+dx^5\wedge dx^6\wedge dx^7\wedge dx^8)
\end{equation}
and ($\mathbb{J}$ below is the Kahler 2-form on $\mathbb{R}^8$, not on $\mathbb{CP}^3$)
\begin{equation}
  G_4=\gamma\mathbb{J}\wedge\mathbb{J}\ \ \ ({\rm with}\ \ \mathbb{J}
  =dx^1\wedge dx^2+dx^3\wedge dx^4+dx^5\wedge dx^6+dx^7\wedge dx^8)
\end{equation}
on $\mathbb{R}^8$, one obtains
\begin{equation}
  m_{b0}^2=\frac{\gamma_0^2}{4}\ \ \ {\rm and}\ \ \ m_b^2=3\gamma^2\ ,
\end{equation}
respectively. All quantities with the subscript $0$ are for the case with the
maximally supersymmetric mass deformation, to be compared with the case of our
interest. In each case, expanding the fermion mass term in (\ref{wv-fermion-mass}),
one obtains
\begin{equation}
  m_{f0}=\pm 2\cdot 4!\beta\gamma_0\ ,\ \ m_f=(m_\xi,m_{\Psi^i})\equiv
  (\pm 12\cdot 4!\beta\gamma,\ \pm 4\cdot 4!\beta\gamma)\ ,
\end{equation}
where the last two masses are those for the complex fermion $\xi$ in ${\bf 1}_{\pm 2}$
and the real fermions $\Psi^i$ in ${\bf 6}_0$. Their mass ratio $3$ is what we anticipated
from group theory. In the former case, since the masses for bosons and fermions are
equal due to supersymmetry, $m_{b0}=m_{f0}$ can be used to conveniently fix the normalization
as $\beta=(4\cdot 4!)^{-1}$. Inserting this back, one obtains
\begin{equation}\label{single-mass}
  m_\xi=\sqrt{3}m_b ,\ \ m_{\Psi^i}=\frac{1}{\sqrt{3}}m_b\ .
\end{equation}
These masses happen to satisfy a supertrace constraint
${\rm Str}(M^2)=0$.\footnote{This constraint turns out to hold for general self-dual
flux. From the fermion mass matrix $M_f=\frac{1}{4\cdot 4!}\gamma^{IJKL}G_{IJKL}$ and the relation (\ref{boson-mass}) to the boson mass, one can show ${\rm Str}(M^2)=8m_b^2-{\rm tr}M_f^2=0$.}

Now we explain how one can obtain the same masses from the asymptotic $AdS_4\times S^7$
gravity solution in the large $N$ limit. Contrary to the simple case with single M2-branes,
which is described by a free quantum field theory, the masses that we calculate below
with the large $N$ gravity solution is the bare mass, or the UV mass, of the field theory.
Since the mass-deformed field theory preserves no supersymmetry, there can be nontrivial
quantum corrections for the actual mass of particles
in the IR. The latter masses will be important quantities when one discusses the
non-relativistic limit from the field theory. The procedure of obtaining fermionic masses  \cite{Bena:2000zb} is similar to the above calculation for single M2-branes.
The 4-form $T_4$ in $\mathbb{R}^8$ defined by $4T_4=dS_3$ and $G_4\sim d\left[\frac{R^6}{r^6}S_3\right]$ plays the role similar to the constant 4-form
which appeared in the mass matrix for M2-branes in flat space, where $R$ is the radius
of 7-sphere which we set $1$ in our normalization. $G_4$ above is the flux on the $S^7$
and the radial part only. Working with the asymptotic solution of the form,
\begin{equation}\label{UV-coordinate}
  ds_{11}^2\sim r^4\left(\eta_{\mu\nu}dx^\mu dx^\nu\right)+\frac{dr^2}{r^2}+ds^2(S^7)
  \ ,\ \ G_4\sim d\left(r^6dt\wedge dx^1\wedge dx^2\right)+d\left(r^{-6}S_3\right)\ ,
\end{equation}
one obtains $T_4=\sqrt{\frac{16a_1}{3}}(dx^{1234}+dx^{1256}+dx^{1278}
+dx^{3456}+dx^{3478}+dx^{5678})$. The resulting fermion masses are
\begin{equation}
  m_\xi=\sqrt{12a_1}\ ,\ \ m_{\Psi^i}=\sqrt{\frac{4a_1}{3}}\ .
\end{equation}
As for the overall normalization, we matched with the convention used in
\cite{Cheon:2011gv} for the maximally supersymmetric mass deformation, in which
$T_4=2\mu_0(dx^{1234}+dx^{5678})$ from the gravity solution leads to the
boson/fermion masses $\mu_0$.\footnote{This relation to mass is obtained by comparing
some BPS spectrum of particles or domain walls on both gauge/gravity sides. The relation
corrects a minor typo of factors in v1 of \cite{Cheon:2011gv}, as explained in
\cite{Hashimoto:2011nn}.} The bosonic mass can be quickly computed if one
takes a probe M2-brane in the mass-deformed UV solution. Using the radial coordinate $r$
appearing in (\ref{UV-coordinate}), one obtains a quadratic potential in $r$ after
partly canceling the contributions from the Nambu-Goto and Wess-Zumino terms. In the
maximally supersymmetric case and for our case, we obtain $m_{b0}=\mu_0$ and
$m_b=2\sqrt{a_1}$, respectively, consistent with (\ref{single-mass}).

Suppose we have a system with particles having various values of mass and
global $U(1)_b$ charge. We consider non-relativistic limit with the $U(1)_b$ symmetry
as we did in our gravity solution. The bosonic and two fermionic fields have charges
$R_b=1$, $2$, $0$, respectively from (\ref{skew-whiffed-rep}).
There could be other particles coming from bound states carrying nonzero global charge.
We take the non-relativistic energy to be $E_{NR}=E_{rel}-aR_b\geq 0$, where $E_{rel}$
is the relativistic energy and $a$ is some constant. The constant $a$ has to be chosen
such that the condition $E_{NR}\geq 0$ is met for all particles, while being
saturated by a subset of them. From the gravity solution, as we made a coordinate
transformation $x^-=\psi-a_0t$, $x^+=t$, the conserved charge for $x^+$ is given by
$E_{NR}=E_{rel}-a_0R_b$, with $a=a_0$. The non-relativisitic field theory only keeps
a sector of particles which satisfy $E_{NR}=0$ when they are at rest.
On the other hand, it discards at low energy all the other particles with
$E_{NR}>0$ at rest, as such particles have large non-relativistic energies.
This is exactly same as discarding anti-particles with non-relativistic energy
$E_{NR}=2mc^2$. So we have $\frac{R_b}{m}\leq a^{-1}$ for some constant $a$,
where the modes which saturate the inequality survive the non-relativistic limit.
Therefore, the non-relativistic theory would be keeping the particles with the
largest value of $\frac{R_b}{m}$, using the IR masses of the particles.

Let us start from the theory on a single M2-brane, described by a free QFT.
Considering the three elementary particles only, one obtains the ratio
\begin{equation}
  \frac{1}{m_b}:\frac{2}{m_\xi}:\frac{0}{m_{\Psi^i}}=1:\frac{2}{\sqrt{3}}:0\ .
\end{equation}
Thus, on single M2-branes, the $\xi$ mode survives in the non-relativistic theory
if one uses $U(1)_b$ as the particle number charge. For the case with large $N$,
had all the UV masses been the physical masses of elementary
particles without any quantum corrections, one would have obtained the same ratio
as the single M2-brane. The actual values for $m/R_b$ from our gravity solution are $\frac{m_\xi}{2}=\sqrt{\frac{9}{8}}a_0$, $m_b=\sqrt{\frac{3}{2}}a_0$ from (\ref{UV-modes})
and (\ref{UV-IR-relation}), which are both larger than $a=a_0$ which we obtained
from the coordinate transformation. There could be two possible reasons for this.
It could be that the physical masses for these particles in IR acquire nontrivial
quantum corrections causing smaller mass-to-charge ratios: at least nonzero quantum
corrections are natural for our non-supersymmetric theories. And/or, there can also
appear bound states of these particles with nonzero binding energies, lowering the ratio.
As the Schr\"{o}dinger geometry comes with an infinite tower of Kaluza-Klein states on $x^-$,
representing infinitely many particle species \cite{Barbon:2009az,Balasubramanian:2010uw},
the last issue of bound states should anyway be an essential ingredient for understanding
this system. It should be interesting to have a better understanding on these quantum
aspects of the spectrum.

\section{Discussions}

In this paper, we considered the gravity solutions for a class of mass-deformed
CFT's in 3 dimensions, and showed that there exist 5 dimensional Schr\"{o}dinger
geometries in the IR after taking the non-relativistic limit with a global $U(1)$
symmetry. As far as we are aware of, this is the first occasion in which 
non-relativistic conformal geometries are explicitly derived by taking the natural
non-relativistic limit of massive relativistic systems. One could think of some
generalizations and extensions of this idea to various directions.

Firstly, one may ask if a similar story holds for the maximally supersymmetric
mass deformation of the same UV theory. As the 4-form flux preserves $SO(4)\times SO(4)$,
the flow solution would have two round 3-spheres. We should be generalizing
\cite{Ooguri:2009cv,Jeong:2009aa} and seek for a solution with broken supersymmetry
if the Witten index calculation of \cite{Kim:2010mr} is implying that the vacuum
spontaneously breaks supersymmetry. Let us try to see if a mechanism explained in this
paper could be working, namely, if the `particle number circle' can shrink in 
IR to be part of the 5 dimensional Schr\"{o}dinger geometry. As the $U(1)$ symmetry
is taken from the two round 3-spheres in \cite{Nakayama:2009cz,Lee:2009mm}, any shrinking
circle necessarily demands that the whole 3-spheres shrink as well, presumably making
the IR geometry very singular. It may also be worth considering the possibility
that this vacuum could be metastable, or does not exist at all.

As explained before, the nonrelativistic CFT of \cite{Nakayama:2009cz,Lee:2009mm}
is based on the `symmetric vacuum' of the supersymmetric mass-deformed theory,
with zero scalar VEV. The vacuum preserves supersymmetry only when $N\!\leq\!k$. 
Its gravity dual will be highly curved, as the 't Hooft coupling is small.
Still, formal gravity solutions with large curvature are identified in
\cite{Cheon:2011gv}. It may be possible to learn something useful from these solutions.

Let us also mention at this point that, from the viewpoint of the non-relativistic limit
of mass-deformed CFT, the appearance of Schr\"{o}dinger geometries as the gravity duals
of non-relativistic CFT's seems to demand a strong restriction. To illustrate
this point, note that there are many ways of obtaining non-relativistic theories with
a given mass-deformed CFT if there exist more than one global $U(1)$ symmetries, simply
by choosing different $U(1)$'s as particle number symmetries. Indeed, for the maximally
supersymmetric mass deformation, different non-relativistic CFT's obtained by choosing
different $U(1)$'s in the $SO(4)\times SO(4)$ global symmetry were studied
\cite{Nakayama:2009cz}. In our mass deformed theory, one can also consider
using a $U(1)$ in the $SU(4)$ isometry as the particle number symmetry, rather than
$U(1)_b$ along the $\psi$ or $x^-$ fiber of $SE_7$ as we did in this paper. Note that
a crucial ingredient which allowed Sch$_5$ of this paper was the shrinking circle for
$U(1)_b$ symmetry, $e^V\rightarrow 0$, in IR. This cannot happen for $U(1)\!\subset\!SU(4)$
as the $KE_6$ part of the metric is finite in IR. This seems to be implying that, for the
geometry of \cite{Son:2008ye,Balasubramanian:2008dm} to emerge as the gravity dual of
Schr\"{o}dinger invariant QFT obtained from mass-deformed relativistic CFT, there
should be nontrivial conditions on the quantum dynamics. It would be interesting
to further explore this issue.

Our solutions reduced down to 4 dimensional spacetime of the gauged supergravity
may not look so well-behaved \cite{Gauntlett:2009bh}. It should be interesting to get a low
dimensional intuition for the solutions with Schr\"{o}dinger symmetry to explore a more general
class of such solutions, maybe using attractor mechanisms \cite{Halmagyi:2011xh}.
Similar studies have been done in \cite{Balasubramanian:2010uw}, which attempts to realize
the particle number symmetry without using internal isometries.

Following the idea of this paper, it might be possible to find flow solutions
from AdS$_5$ to Sch$_6$ by studying the consistent Kaluza-Klein truncation of
\cite{Cassani:2010uw} based on $AdS_5\times S^5$.
There also exists a holographic flow solution for the 4 dimensional $\mathcal{N}\!=\!4$
Yang-Mills theory with an $\mathcal{N}\!=\!1$ supersymmetric mass deformation, which is
singular in the IR \cite{Girardello:1999bd}.

Perhaps it might be worth mentioning that the original motivation in
\cite{Son:2008ye,Balasubramanian:2008dm} of using this geometry to holographically study
cold fermionic atoms at unitarity may be demanding alternative realizations of the
Schr\"{o}dinger symmetry without using the isometry of $x^-$ as the particle number. This
question has been raised due to an exotic thermodynamics of Schr\"{o}dinger black holes
\cite{Herzog:2008wg}, probably caused by the presence of an infinite tower of
Kaluza-Klein particle species \cite{Barbon:2009az,Balasubramanian:2010uw}.
Of course our system would be showing same phenomena, simply due to the appearance of the
geometry of the form (\ref{schrodinger}). The non-relativistic limit of \cite{Nakayama:2009cz,Lee:2009mm} is based on using the D0-brane charge (geometrizable
to M-theory isometry) as the particle number, which is also the case for our system.
The non-relativistic CFT discussed in \cite{Herzog:2008wg}
essentially has this $x^-$ direction as it is obtained from a $3+1$ dimensional QFT in
the UV, where the $x^-$ isometry simply comes from the extra direction of the UV field
theory. On the other hand, non-relativistic QFT's obtained by the standard non-relativistic
limit of mass-deformed QFT do not necessarily have such an infinite tower of bound
states. From the field theory perspective, the special example of M2-brane CFT discussed
in this paper seems to be rather exceptional, containing geometrizable D0-brane global
charge and exhibiting bound states for all D0-brane numbers. Thus, it could be possible
to seek for a study along the line of \cite{Balasubramanian:2010uw}, to have a gravity
solution invariant under Schr\"{o}dinger symmetry with non-geometrizable particle number
symmetry in the context of the mass-deformed CFT.

We would also like to comment that in \cite{Gauntlett:2009bh}, the phases of skew-whiffed
field theory with relevant deformations (i.e. coupled to nonzero $h$ field in the gravity
dual) have been studied extensively from the gravity dual, after considering more
modes than what we did in this paper.\footnote{The comments below are motivated by
discussions with Jerome Gauntlett and Julian Sonner.} More precisely, a charged scalar
and a massless gauge field were kept, apart from $h$. In particular, there is a phase
transition which appears at nonzero chemical potential $\mu$ and mass, the latter one being
the mass that we considered in this paper. When $\mu$ is sufficiently larger than the mass,
the system is shown to be in a superconducting phase with spontaneously broken $U(1)$
symmetry. On the other hand, if the chemical potential is small compared to the mass, the
system is supposed to be in a normal phase which is gapped, without the Goldstone boson
associated with spontaneously broken $U(1)$. In particular, the line $\mu=0$
at zero temperature on the parameter space that we have considered all belongs to this
normal phase. At zero temperature in the normal phase, with all the propagating degrees
being massive, one may ask if there are interesting non-relativistic solutions like what
we considered on the $\mu=0$ line. Although the numerical solution was explained to
be singular at zero temperature, this could be physically acceptable. In string theory, there
are examples of four-dimensional extremal black holes described by singular geometries, yet
being supersymmetric \cite{Garfinkle}. Furthermore if we are interested in
condensed matter systems with unique ground states, the dual geometries have zero entropy,
which often lead to singular geometries.  Also, uplifting the 4 dimensional geometry to 11
dimensions could reveal more interesting structures, as was the case on the $\mu=0$ line
we studied in this paper.

Finally, it could be desirable to investigate the stability of the solutions
discussed in this paper. An instability was reported for a solution in the
skew-whiffed truncation with nonzero values of the charged field \cite{Bobev:2010ib}.
Although the solution we discussed is more complicated than the AdS solution
studied there, at least studies of the IR solution (\ref{IR-solution}) would be doable
and worthwhile. Perhaps this problem could be related to the stability analysis of a
class of $AdS_5\times KE_6$ solutions \cite{Martin:2008pf}.

\vskip 0.5cm  \hspace*{-0.8cm}
{\bf\large Acknowledgements}
\vskip 0.2cm

\hspace*{-0.75cm} We would like to thank Aleksey Cherman, Aristomenis Donos, Shamit Kachru,
Sangmin Lee, Sungjay Lee, Eoin 'O Colgain, Sandip Trivedi, Daniel Waldram, Toby Wiseman,
Jackson Wu and especially Jerome Gauntlett and Julian Sonner for helpful discussions.
This work is supported by the BK21 program of the Ministry of Education, Science and
Technology (HK, SK), KOSEF Grant R01-2008-000-20370-0 (JP), the National Research Foundation
of Korea Grants No. 2006-0093850 (KL), 2009-0084601 (KL), 2009-0085995 (JP), 2010-0007512 (HK, SK), 2009-0072755,
2009-0084601 (HK) and 2005-0049409 through the Center for Quantum Spacetime (CQUeST) of Sogang
University (KL, JP). SK thanks the theory group of Imperial College London and DAMTP for
hospitality during his visit, where this work was completed. JP appreciates APCTP for its
stimulating environment for research.

\end{document}